\documentclass[twocolumn,aps,prl,showpacs]{revtex4}
\usepackage{mathptmx}
\usepackage{xspace}
\usepackage{amssymb}
\usepackage{graphicx}

\newcommand{\mub}{$\mu_\mathrm{B}$\xspace}
\newcommand{\tc}{$T_{\mathrm{c}}$\xspace}
\newcommand{\ngd}{$N_{\mathrm{Gd}}$}
\newcommand{\pgd}{$p_{\mathrm{Gd}}$\xspace}
\newcommand{\pef}{$p_{\mathrm{eff}}$\xspace}
\newcommand{\pmt}{$p_{\mathrm{0}}$\xspace}

\newcommand{\celsius}{$^{\circ}$C\xspace}

\newcommand{\etal}{\emph{et al.}}
\newcommand{\jap}{J.\ Appl.\ Phys.\xspace}

\begin{document}

\title{GaN:Gd: A superdilute ferromagnetic semiconductor with a Curie temperature above 300 K}
\author{S.\ Dhar}
\email{dhar@pdi-berlin.de}
\author{O.\ Brandt}
\author{M.\ Ramsteiner}
\author{V.\ F.\ Sapega}
\altaffiliation[Permanent address:]{Ioffe Physico-Technical Institute, Russian Academy of Sciences, 194021 St. Petersburg, Russia}
\author{K.\ H.\ Ploog}
\affiliation{Paul-Drude-Institut f\"ur Festk\"orperelektronik,
Hausvogteiplatz 5--7, D-10117 Berlin, Germany}

\begin{abstract}
We investigate the magnetic and magneto-optic properties of epitaxial GaN:Gd layers as a function of the external magnetic field and temperature. An unprecedented magnetic moment is observed in this diluted magnetic semiconductor. The average value of the moment per Gd atom is found to be as high as 4000~\mub as compared to its atomic moment of 8~\mub. The long-range spin-polarization of the GaN matrix by Gd is also reflected in the circular polarization of magneto-photoluminescence measurements. Moreover, the materials system is found to be ferromagnetic above room temperature in the entire concentration range under investigation (7$\times$10$^{15}$ to 2$\times$10$^{19}$~cm$^{-3}$). We propose a phenomenological model to understand the macroscopic magnetic behavior of the system. Our study reveals a close connection between the observed ferromagnetism and the colossal magnetic moment of Gd.
\end{abstract}

\maketitle
Efficient spin injection into semiconductor structures is one of the most important requirements to realize future spintronic devices. A ferromagnetic material compatible with semiconductor growth which thus could easily be integrated into semiconductor devices would be the best choice. Since the discovery of a Curie temperature T$_C$ as high as 110~K in (Ga,Mn)As \cite{ohno}, III-V dilute magnetic semiconductors (DMS) have gained a lot of attention. GaN-based DMS have the potential for exhibiting high temperature ferromagnetism which is essential for developing future spintronic devices \cite{dietl}. Several transition metal (TM) doped GaN DMS have been investigated. Room-temperature ferromagnetism in these materials, however, is often found to result from the formation of precipitates rather than from a homogeneous alloy \cite{dhar}. The rare-earth (RE) elements could be an interesting alternative to transition metals. RE atoms have partially filled $f$ orbitals which carry magnetic moments and may take part in magnetic coupling like in the case of TM with partially filled $d$ orbitals. While it is expected that the magnetic coupling strength of $f$ orbitals is much weaker than that of $d$ orbitals due to the stronger localization of the $f$ electrons, Gd is the only RE element which has both partially filled 4$f$ and 5$d$ orbitals. Together these orbitals can take part in a new coupling mechanism proceeding via intra-ion 4$f$-5$d$ exchange interaction followed by inter-ion 5$d$-5$d$ coupling mediated by charge carriers \cite{story}.

Here, we perform a systematic study of the magnetic properties of the DMS GaN:Gd. The magnetization measurements reveal an unprecedented magnetic moment of up to 4000~\mub per Gd atom. This colossal magnetic moment can be explained in terms of a long-range spin-polarization of the GaN matrix by Gd. This conclusion is independently confirmed by the pronounced changes of the circular polarization in magneto-photoluminescence measurements upon light Gd-doping.  Moreover, all samples are found to be ferromagnetic above room temperature. A phenomenological model is developed to explain the macroscopic magnetic behavior of the system,  suggesting a close connection between the observed ferromagnetism and the colossal magnetic moment of Gd.  

The 400--700~nm thick GaN layers with a Gd concentration ranging from 7$\times$10$^{15}$ to 2$\times$10$^{19}$~cm$^{-3}$ were grown directly on 6H-SiC(0001) substrates using reactive molecular-beam epitaxy (RMBE). The RMBE system (base pressure: 2$\times$10$^{-10}$~Torr) is equipped with conventional effusion cells for Ga (purity: 7N) and Gd (purity: 4N) \cite{remark} and an unheated NH$_3$ (purity: 7N) gas injector. A substrate temperature of 810\celsius (our standard growth temperature for GaN) was used. The Gd/Ga flux ratio was changed in order to adjust the Gd concentration in the layers. Nucleation and growth was monitored \textit{in situ} by reflection high-energy electron diffraction (RHEED). A spotty (1$\times$1) pattern, reflecting a purely three-dimensional growth mode, was observed during nucleation of the layers. The pattern quickly became streaky, reflecting two-dimensional growth.

The Gd concentration of the layers was determined by secondary ion mass spectrometry using a CAMECA IMS 4f system, employing 0$_2^+$ primary ions with an impact energy of 7~keV. Mass interference due to parasitic molecular ions was avoided by selecting precisely the masses of two Gd isotopes (155.922 and 157.924). The resulting ion rates were identical to within 1\%. Measurements at different locations, each of which involved an area of 50$\times$50~$\mu$m$^{2}$, deviated by no more than 5\%. The concentration was calculated according to measurements of an ion-implanted standard (sample I), which was implanted with a Gd dose of 10$^{15}$~cm$^{-2}$. The detection limit was found to be 2$\times$10$^{15}$~cm$^{-3}$.  

Magnetization measurements up to 360~K were done in a Quantum Design superconducting quantum interference device (SQUID) magnetometer with the sample held in a plastic straw provided by Quantum Design. The response of the magnetometer was calibrated with epitaxial Fe and MnAs layers on GaAs for which the magnetization is accurately known. The magnetization of undoped GaN layers was found to be indistinguishable from that of bare SiC substrates. Magnetization loops were recorded at various temperatures for magnetic fields between $\pm$50~kOe. The magnetic field was applied parallel to the sample surface, i.\ e., perpendicular to the c-axis, for all measurements. All data presented here were corrected for the diamagnetic background of the substrate.

For magneto-PL, the sample was excited by a He-Cd ion laser at a wavelength of
325~nm. The laser power densities focused on the sample were in the range
between 200 and 250~Wcm$^{-2}$. The experiments were carried out in the temperature range 5--100~K, in a continuous He-flow cryostat, and in magnetic fields up to 12~T using the
backscattering Faraday geometry. In all measurements, the magnetic field 
was oriented parallel to the c-axis of the samples. The magnetic field 
induced circular polarization of the (D$^0$,X) transition was analyzed by passing 
the emitted light through a photoelastic modulator (PEM), a quarter-wave-plate, 
and a linear polarizer.

\begin{figure}[t!]
\centerline{\includegraphics*[width=7.0cm]{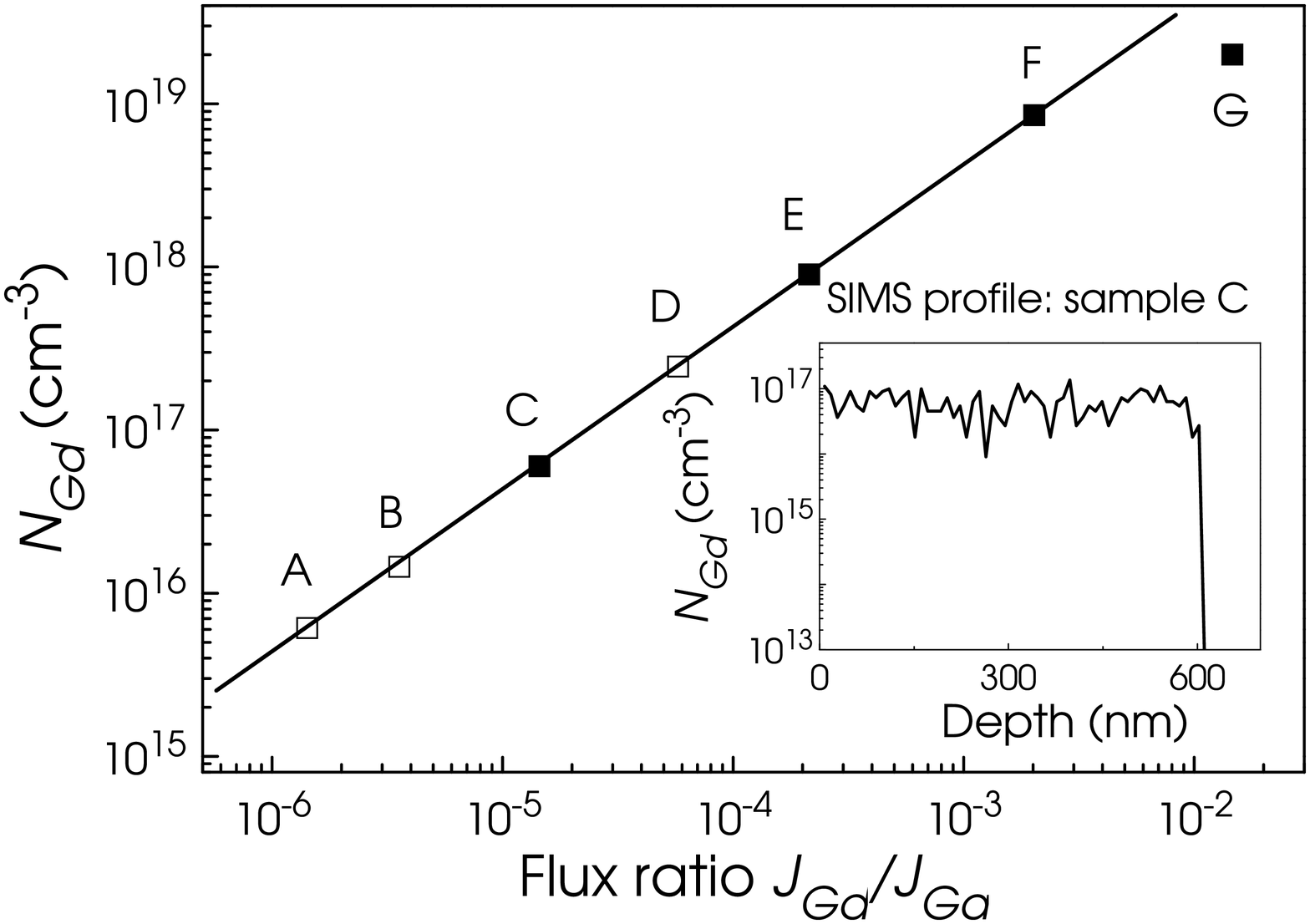}}
\caption{Gd concentration as measured by SIMS as a function of Gd/Ga flux ratio (solid squares). The solid line is a linear fit to the data representing sample C, E and F. The Gd concentration for sample A, B and D (open squares) is obtained from the corresponding  Gd/Ga flux ratio by extrapolation. The inset of the figure shows the SIMS profile obtained for sample C.}
\label{fluxr}
\end{figure}

GaN:Gd layers A--G were grown with the Gd/Ga flux ratio increasing from A to G in alphabetic order. The concentration of Gd (\ngd) in samples C, E, F and G is measured by secondary ion mass spectrometry (SIMS). In Fig.\ \ref{fluxr}, \ngd\ as measured by SIMS is plotted as a function of the Gd/Ga flux ratio $\phi$. Samples C, E and F (solid squares) are lying on a straight line with a slope of unity indicating a linear dependence of Gd incorporation on $\phi$. The Gd concentration of sample G is smaller than expected from this linear dependence. In fact, we have observed a strongly faceted surface for sample G, indicating that Gd is affecting the growth mode at these high concentrations. \ngd\ for sample A, B and D (open squares) is obtained by linearly extrapolating the curve passing through samples C, E and F. The inset of the figure shows the SIMS profile obtained for sample C. The concentration of Gd is found to be constant over the entire depth for sample C and all other samples investigated by SIMS, ruling out any accumulation of Gd on the surface during growth. All samples were subject to an extensive investigation by high-resolution x-ray diffraction in a wide angular range. No reflections related to a secondary phase was detected. Furthermore, we have investigated one sample (C) by transmission electron microscopy. No clusters or precipitates were observed. Electrically, all Gd doped samples are found to be highly resistive.

\begin{figure}[t!]
\centerline{\includegraphics*[width=7.0cm]{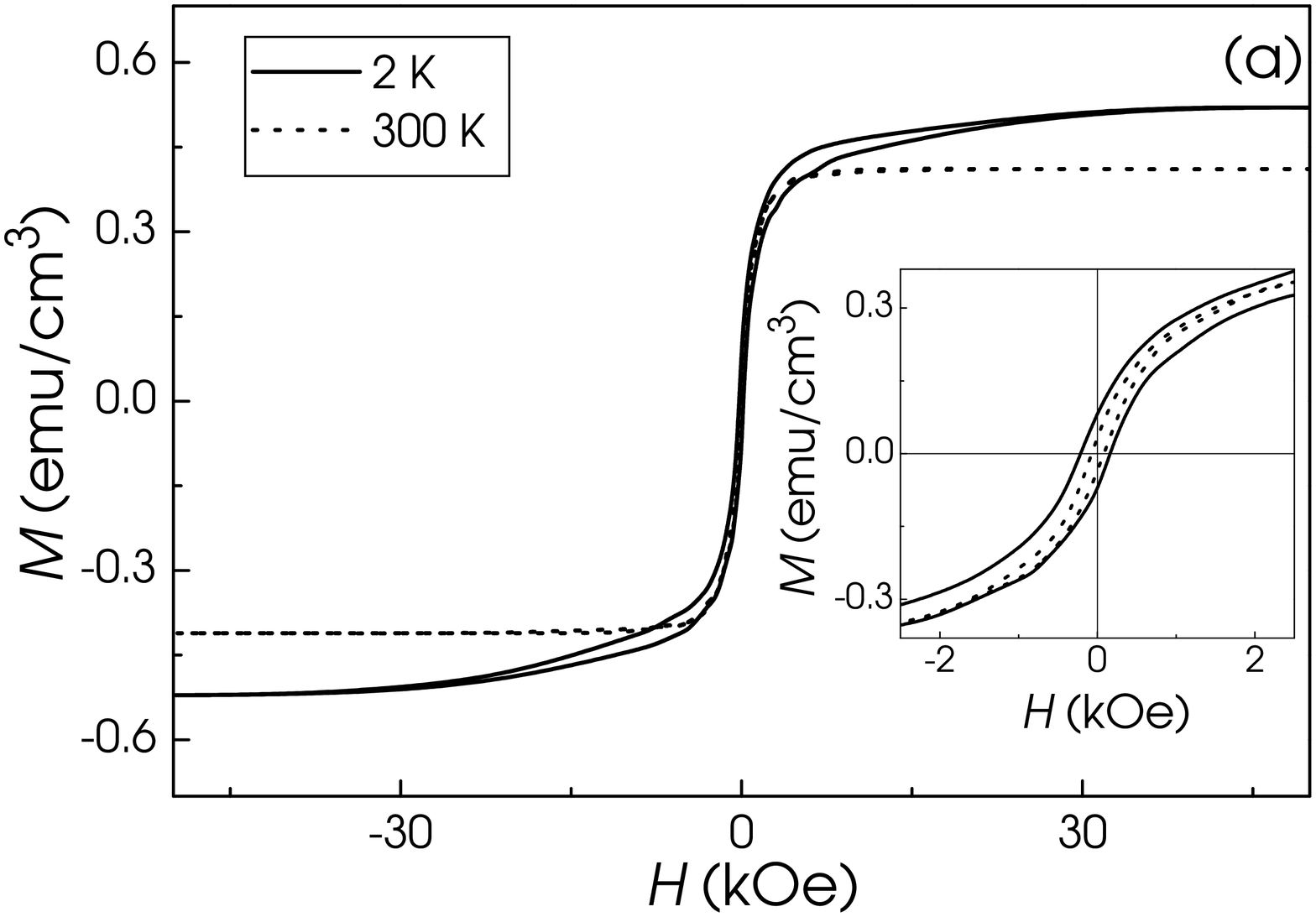}}
\centerline{\includegraphics*[width=7.0cm]{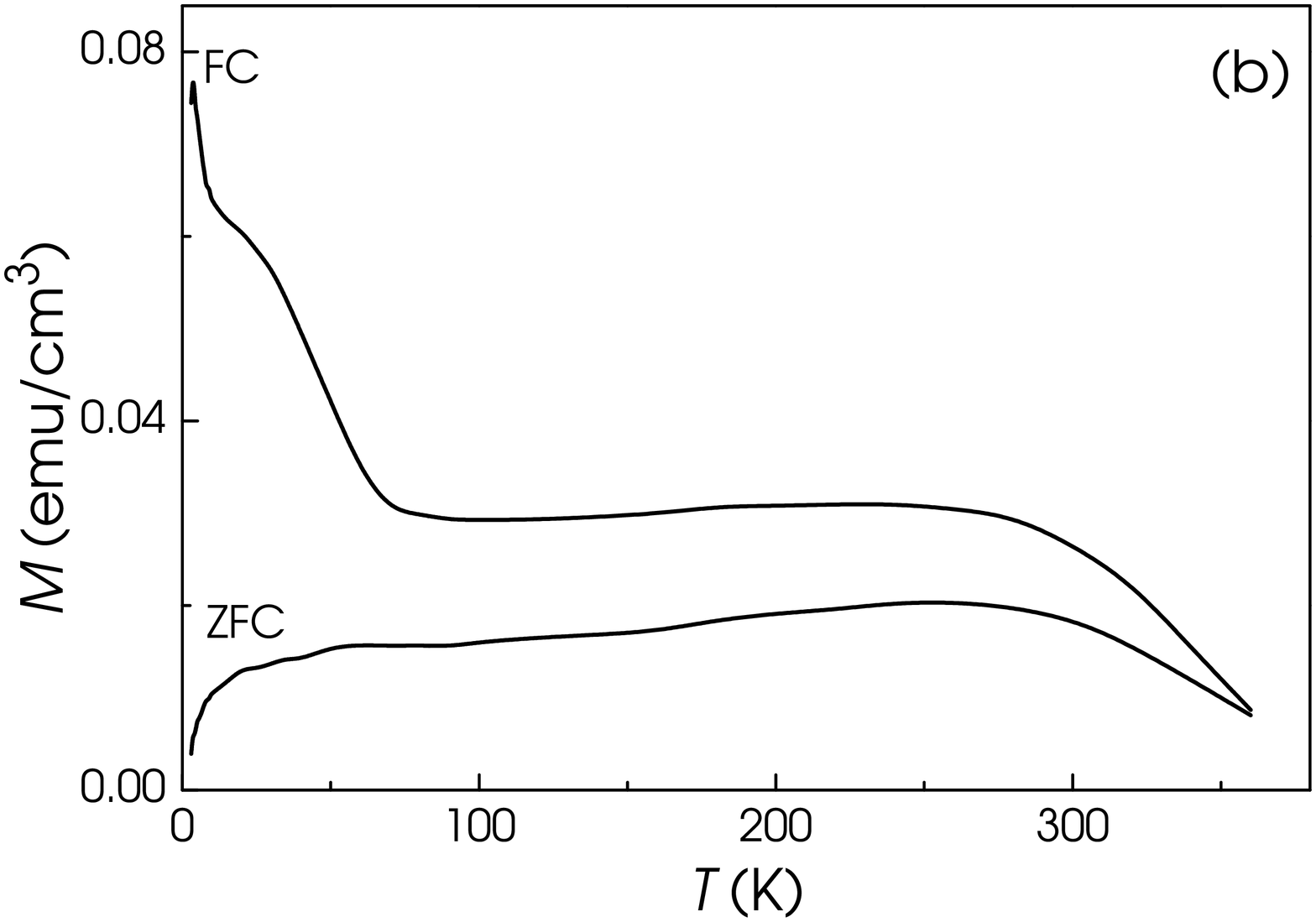}}
\centerline{\includegraphics*[width=7.0cm]{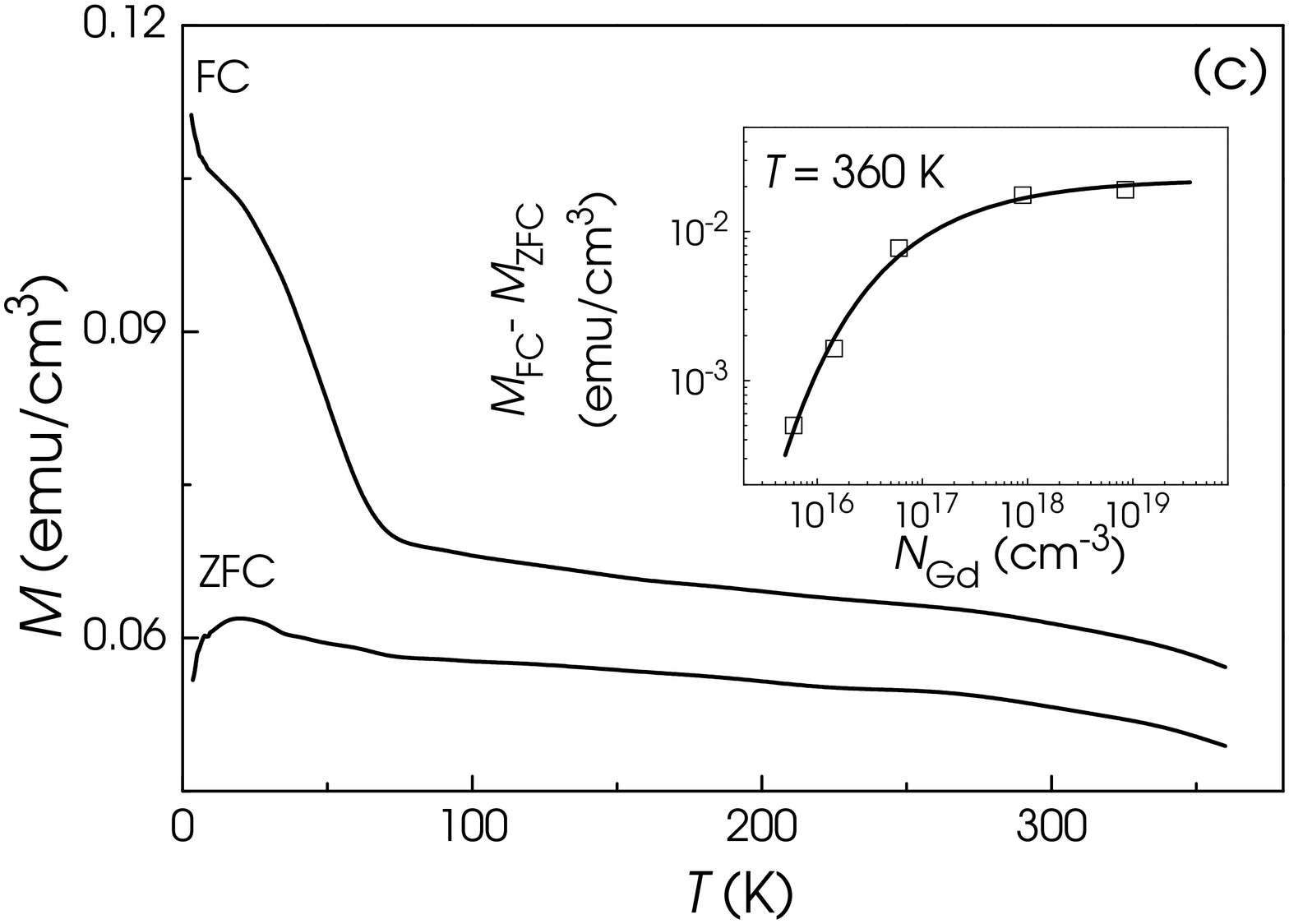}}
\caption{(a) Magnetization loops obtained for sample C at 2~K (solid line) and 300~K (dotted line). The inset of (a) shows the loops at small fields. Temperature dependence of magnetization at field-cooled (FC) and zero-field-cooled (ZFC) conditions at a magnetic field of 100~Oe for (b) sample A  and (c) sample C. The difference between FC and ZFC magnetization measured at 360 K is plotted as a function of \ngd\ in the inset of (c). The solid curve through the data is guide to the eye.}
\label{fczfchys}
\end{figure}

Figure \ref{fczfchys} (a) shows the magnetization loops obtained for sample C at 2~K and 300~K. At both temperatures, the magnetization saturates at high magnetic fields (unlike superparamagnetic materials) and exhibits a clear hysteresis at lower fields. These two features clearly indicate a ferromagnetic behavior. We have observed qualitatively similar hysteresis loops for all samples. Figure \ref{fczfchys} (b) and (c) show the temperature dependence of the magnetization under a magnetic field of 100~Oe for samples A and C, respectively. Prior to measuring the temperature dependence of the magnetization, the sample is first cooled from room temperature to 2~K either under a saturation field of 20~kOe (Field cooled: FC) or at zero field (zero field cooled: ZFC). It is clear from these figures that the FC and ZFC curves are qualitatively similar for both samples featuring a double step like structure below 100~K in the FC curves. In the case of sample C, the two curves remain separated throughout the entire temperature range (2--360~K),  while they coincide at around 360~K for sample A.
The separation between the FC and ZFC curves indicates a hysteretic behavior which is consistent with our previous observation from Fig.\ \ref{fczfchys} (a). The two curves coincide at the Curie temperature \tc, when the hysteresis disappears. Clearly, \tc is around 360~K for sample A while it is much larger for sample C.  The separation between the FC and ZFC curves measured at 360~K is plotted as a function of Gd concentration in the inset of Fig.\ \ref{fczfchys} (c), and is seen to increase with increasing Gd concentration revealing a shift of \tc towards higher temperatures. We have also investigated a Gd-implanted GaN sample (sample I) with a higher Gd concentration than we could reach by incorporation during growth. This sample also exhibits ferromagnetism at and well above room-temperature. These observations are consistent with the results obtained by Teraguchi \etal, who observed a Curie temperature larger than 400~K in Ga$_{0.94}$Gd$_{0.06}$N alloys \cite{teraguchi}. 

\begin{figure}[t!]
\centerline{\includegraphics*[width=7.0cm]{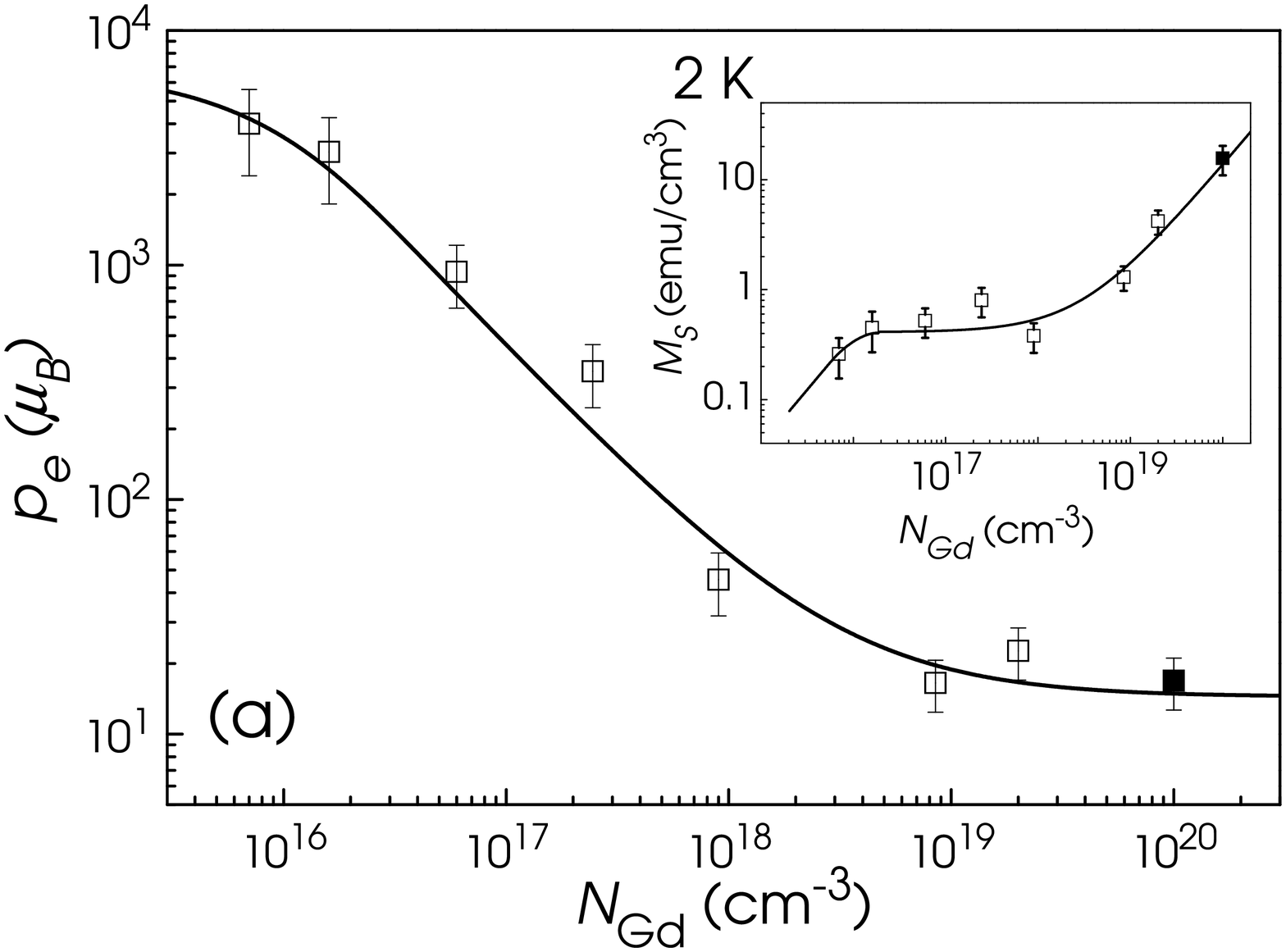}}
\centerline{\includegraphics*[width=7.0cm]{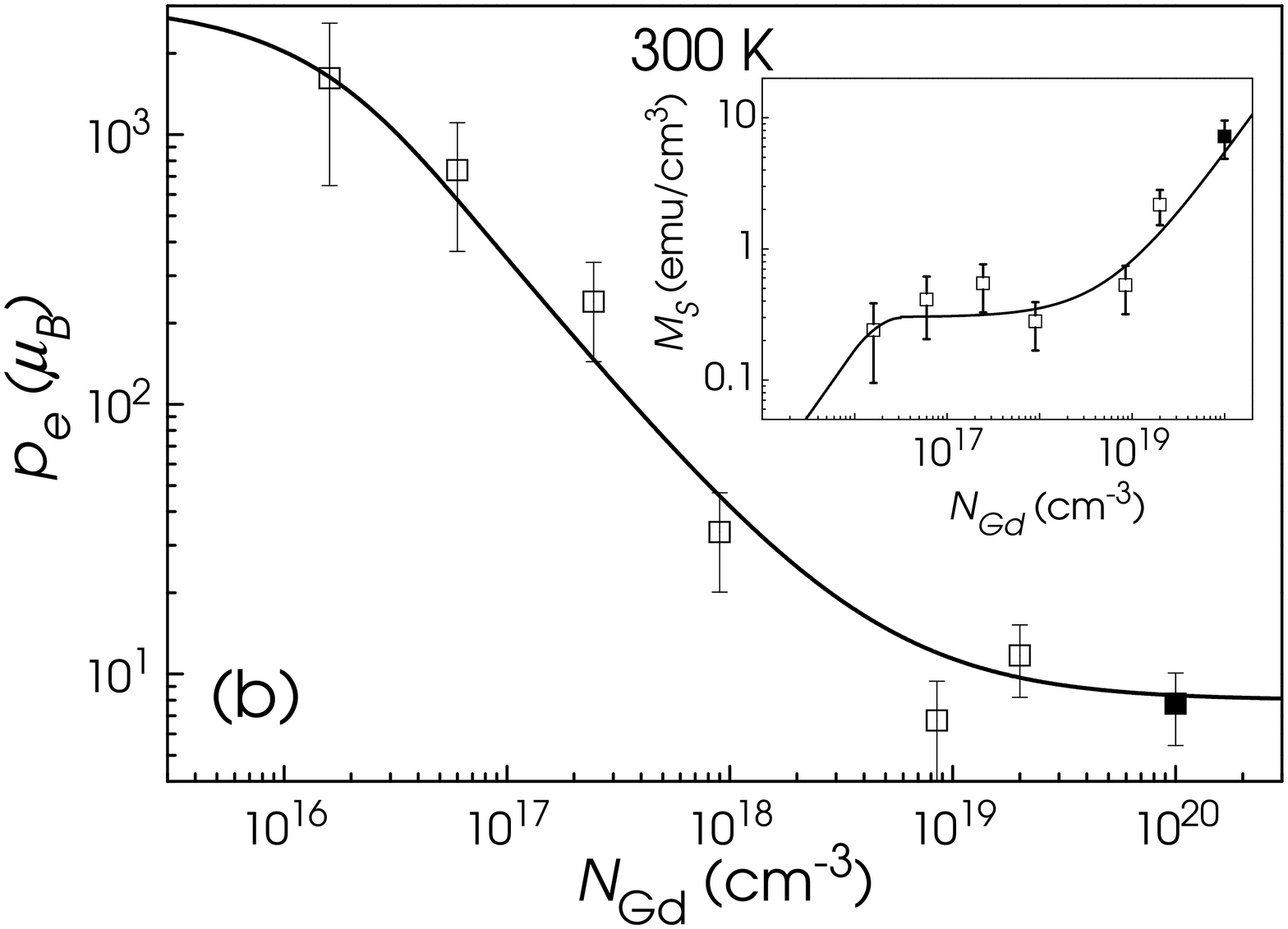}}
\caption{Magnetic moment per Gd atom (\pef) as a function of Gd concentration at (a) 2~K and (b) 300~K. The saturation magnetization ($M_s$) as a function of Gd concentration at 2 and 300~K is also shown in the respective insets. Open squares are representing the experimental data for sample A--G and solid squares for sample I. Solid lines are the theoretical fit obtained from our model (discussed in the text).}
\label{peffn}
\end{figure}

The effective magnetic moment per Gd atom \pef can be obtained from the value of the saturation magnetization $M_s$ (\pef = $M_s$/\ngd). In sample C, \pef is found to be 935 and 737~\mub at 2~K and 300~K, respectively. These values are about two order of magnitudes larger than the pure moment of Gd. In Fig.\ \ref{peffn}, \pef obtained at different temperatures is plotted as a function of \ngd.  Data obtained from sample I are also included in this figure (solid squares). Both at 2 K [Fig.\ \ref{peffn} (a)] and 300~K  [Fig.\ \ref{peffn} (b)], \pef is found to be extraordinarily large, particularly for low Gd concentrations. At high Gd concentrations, \pef is seen to saturate. It is interesting to note that at 300~K, the saturation value is close to that of the atomic moment of Gd (8~\mub), while at 2~K, it is clearly higher. This is true even for sample I. Such a colossal moment can be explained in terms of a very effective spin-polarization of the GaN matrix by the Gd atoms. In fact, in certain diluted metallic alloys, the solute atoms exhibit an average magnetic moment larger than their atomic value. This effect is called giant magnetic moment. There are several reports on a giant magnetic moment of Fe, Mn and Co when they are either dissolved in or residing on the surface of Pd \cite{bergmann,flouquet,crangle}. This effect is explained in terms of a spin polarization of the surrounding Pd atoms. Very recently, a giant magnetic moment was observed in Co doped SnO$_{2-\delta}$ \cite{ogale}, a DMS. High-temperature ferromagnetism and an unusually large magnetic moment were observed also in Fe-doped SnO$_2$, which was argued to originate from ferrimagnetic coupling mediated by electrons trapped on bridging F-centers \cite{coey}.

\begin{figure}[t!]
\centerline{\includegraphics*[width=7.0cm]{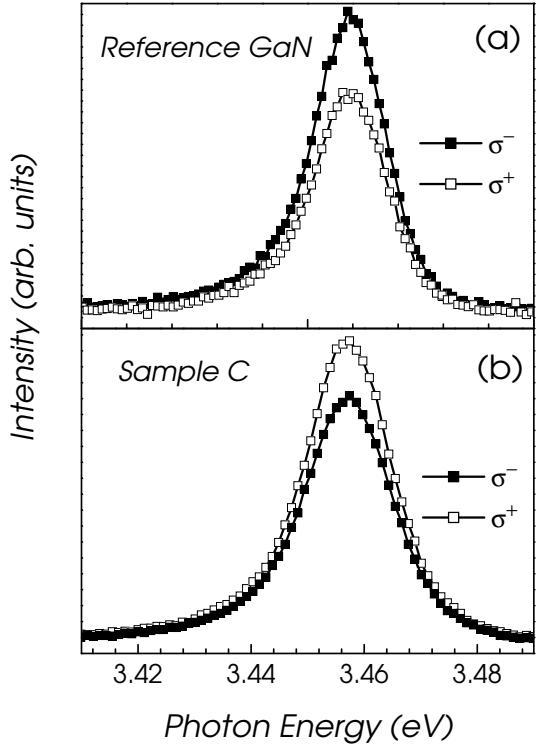}}
\caption{Circularly polarized PL spectra for (a) the reference GaN sample and (b) sample C, measured at 6~K under a magnetic field of $B$ = 10~T in Faraday configuration ($B$ $\|$ $c$-axis).}   
\label{plpol}
\end{figure}

The above conclusion of a long-range spin-polarization of the GaN matrix by the Gd atoms is independently confirmed by magneto-PL. In the absence of an external magnetic field, the PL spectra of all samples are characteristic for undoped epitaxial GaN layers in that they are dominated by a neutral-donor bound exciton (D$^0$,X) transition at $\sim$3.458~eV. The lower energy of this transition when compared to homoepitaxial GaN is consistent with the tensile in-plane strain in these layers of 0.15\% \cite{waltereit}. In all cases, the donor responsible for this transition is O with a concentration of about 10$^{18}$~cm$^{-3}$ as measured by SIMS. However, clear differences between undoped and Gd-doped GaN layers are evident when analyzing the circular polarization of the PL in the presence of a magnetic field. Figure \ref{plpol} shows the PL spectra for a GaN reference sample [Fig.\ \ref{plpol} (a)] and sample C [Fig.\ \ref{plpol} (b)] under a magnetic field of $B = 10~$T. The transition is polarized for both samples, which is evident from the difference in intensities of the two circularly polarized $\sigma^+$ (open squares) and $\sigma^-$ (solid squares) components. Most importantly, the polarization has opposite sign for sample C when compared to the reference sample, which clearly indicates that the excitonic ground state in sample C is subjected to an internal magnetic field which is effectively stronger than the external field and is acting in the opposite direction. Considering that the mean Gd-Gd separation in sample C is 25~nm, and that the average distance of a (D$^0$,X) site from a Gd atom will thus be as large as 12~nm \cite{note1}, this finding directly confirms our previous conclusion that the Gd atoms are introducing a long range spin polarization in the GaN matrix. We also observe that the relative change of polarization with respect to the reference sample increases with the Gd concentration, as intuitively expected. These results, together with a quantitative analysis of the dependence of the polarization on the magnetic field, will be reported in detail elsewhere \cite{victor}. 

In order to get a quantitative understanding of the range of the spin polarization, we have developed a phenomenological model as explained in the following. The polarization of the GaN matrix by the randomly positioned Gd atoms is described as a rigid sphere of influence around each Gd atom, meaning that all the matrix atoms within the sphere are polarized by an equal amount whereas matrix atoms falling outside of this sphere are not affected. Let us associate an induced moment of $p_{\mathrm{0}}$ with each of the matrix atoms lying in the region occupied by one sphere. As one should expect an increase of the polarization of matrix atoms if they belong to a region where two (or more) spheres of influence overlap, we attribute an additional moment $n p_1$ to matrix atoms occupying sites in regions where $n$ ($n=2,...N_{\mathrm{Gd}}$) spheres overlap. 
Since their radius $r$ is presumably much larger than the lattice spacing in GaN, we further assume that the spheres are randomly arranged in a three-dimensional continuum (continuum percolation). Within this framework, the saturation magnetization can be expressed as 
\begin{equation}
M_{\mathrm{s}} = p_{\mathrm{Gd}} N_{\mathrm{Gd}}  + p_{\mathrm{0}} \widetilde{v} N_{\mathrm{0}} + p_1 N_0 \sum_{n = 2}^{N_{\mathrm{Gd}}} n \widetilde{v}_n
\end{equation}
\noindent where $N_0$ is the concentration of matrix atoms per unit volume, $v$ is the volume of each sphere, $\widetilde{v} = 1 - \exp(- v N_{\mathrm{Gd}})$ is the volume fraction occupied by the spheres, and 
\begin{equation}
\widetilde{v}_n=\frac{(v N_{\mathrm{Gd}})^n}{n!} e^{- v N_{\mathrm{Gd}}}
\end{equation}
 is the volume fraction of the regions contained within $n$ spheres. The average magnetic moment per Gd atom \pef is then obtained as 
\begin{eqnarray}
p_{\mathrm{eff}} &=& p_{\mathrm{Gd}} + p_1 N_{\mathrm{0}} v \nonumber \\
 &+& \left [ p_{\mathrm{0}} - (p_{\mathrm{0}} + p_1 N_{\mathrm{Gd}} v) e^{- v N_{\mathrm{Gd}}} \right ] \frac{N_{\mathrm{0}}}{N_{\mathrm{Gd}}}
 \label{eq3}
\end{eqnarray}
 
At low Gd concentrations, most of the spheres are well separated and \pef has its maximum value. As the concentration of Gd is increased, more and more spheres overlap and \pef decreases. Finally, at a very high Gd concentration the entire matrix becomes polarized, and \pef approaches saturation. Note that the value of saturation is larger than the magnetic moment of bare Gd atoms by an amount of $p_1 N_{\mathrm{0}} v$. We use Eqn.~\ref{eq3} to fit our experimental data with \pgd = 8~\mub and \pmt, $p_1$ and $r$ as free parameters. The agreement is quite satisfactory as shown in Fig.~\ref{peffn}. The fit yields  \pmt = $1.1 \times 10^{-3}$~\mub, $p_1 = 1.0 \times 10^{-6}$~\mub, and $r = 33$~nm at 2~K and \pmt = $8.4 \times 10^{-4}$~\mub, $p_1 \approx 0$, and $r = 28$~nm at 300~K. The finite value of $p_1$  at 2~K explains why \pef saturates at a value which is still higher than the atomic moment of Gd. Most important, however, is that the value for $r$ is sufficiently large to account for the strong effect of Gd on the excitonic ground state of the GaN matrix (\emph{cf.} Fig.\ \ref{plpol}). 

The second remarkable property of our GaN:Gd layers is the high-temperature ferromagnetism observed in all samples (well above room temperature, compared to a \tc of 289 K for bulk Gd). We believe that the ferromagnetism and the colossal moment of Gd observed in these samples are in fact closely related. An overlap of the spheres of influence establishes a (long-range) coupling between the individual spheres. Within the framework of percolation theory, ferromagnetism is expected to occur at the percolation threshold, when an "infinite cluster" spanning macroscopic regions of the sample is formed. The percolation threshold is reached for our model at $\widetilde{v}$ = 0.28955 \cite{consiglio,baker}. With increasing Gd concentration, we would thus expect a phase transition from paramagnetic to ferromagnetic behavior. Furthermore, \tc which depends upon the strength of the overlap is expected to increase with the Gd concentration. This is clearly consistent with the results shown in the inset of Fig.\ \ref{fczfchys} (c). 

Unfortunately, the well-established percolation formalism cannot be straightforwardly adopted to quantitatively explain the current problem. It is intuitively clear that the spheres of influence are not hard but soft, i.~e., the polarization of the matrix induced by the Gd atoms must decay with increasing distance. The onset of ferromagnetic order now depends upon the precise shape of this polarization cloud as well as on the overlap of two or more of these clouds. A quantitative prediction of the onset of ferromagnetism in this situation thus requires a detailed understanding of the nature of the ferromagnetic coupling in this material.

Regarding the microscopic origin of this coupling, it is clear that it cannot be explained simply in terms of direct, double or superexchange between Gd atoms since the average Gd-Gd distance is too large for such a coupling to exist. Furthermore, all samples are found to be electrically highly resistive, ruling out free-carrier mediated RKKY (Rudermann--Kittel--Kasuya--Yosida) type long-range coupling. 
Finally, also very recent proposals aimed at explaining a giant magnetic moment and/or high-temperature ferromagnetism in various material systems \cite{bergmann,flouquet,crangle,ogale,coey} do not apply to the current requirement of a long-range coupling. An actual understanding of the phenomenon observed in the present paper will require detailed \emph{ab initio} studies, particularly of the $f$-$d$ coupling between Gd and Ga. Considering the long spatial range of the coupling evident from our experiments, such calculations are computationally very demanding and are therefore beyond the scope of the present paper. An important result for any theoretical approach was recently obtained by Hashimoto \etal, who showed that Gd occupies primarily Ga sites in GaN even at a concentration as high as 6\% (two orders of magnitude higher than the highest concentration investigated here) \cite{hashimoto}.  

Our study has revealed an extraordinarily large magnetic moment of Gd in GaN which can be explained in terms of a long-range spin-polarization of the GaN matrix by the Gd atoms. The material is found to exhibit ferromagnetism above room temperature even with a Gd concentration less than 10$^{16}$~cm$^{-3}$. This finding offers an exciting opportunity, since GaN:Gd may be easily doped with donors (acceptors) with a concentration exceeding that of Gd to generate spin polarized electrons (holes) in the conduction band (valence band). Gd-doped GaN with its \tc above room temperature might thus be a very attractive candidate as a source of spin-polarized carriers for future semiconductor-based spintronics \cite{wolf}.

We are indebted to A.\ Trampert, K.~J.\ Friedland and U.\ Jahn for important contributions to this work. Furthermore, we thank J.\ Herfort, M.\ Bowen and R.\ Koch for valuable discussions and suggestions. We also acknowledge partial financial support of this work by the Bundesministerium f\"ur Bildung und Forschung of the Federal Republic of Germany.


\begin{thebibliography}{99}
\bibitem{remark} The purity of the Gd ingot has been determined by the manufacturer (Stanford Materials Corporation) via inductively-coupled plasma mass spectroscopy. The datasheet of our ingot certifies that the ferromagnetic elements Fe, Ni and Co are present in residual traces of $\ll$100~ppm, $\ll$50~ppm, and below the detection limit (0.5~ppm), respectively. Note that the magnetization observed in this work for, e.~g., samples A and C would require an Fe (Ni, Co) concentration of 10$^{19}$~cm$^{-3}$ and above, three orders of magnitude higher than the Gd concentration. Also note that the vapor pressure of all these elements is almost identical to that of Gd.
\bibitem{note1}{It can be easily shown that the average distance of an exciton binding site from its nearest Gd atom is = 0.480296 $a$, where $a$ is the average Gd-Gd separation}
\end{thebibliography}
\end{document}